\documentclass[pre,onecolumn,amsmath,amssymb,superscriptaddress]{revtex4-2}

\usepackage{amsmath}
\usepackage{hyperref}
\usepackage{graphicx}
\usepackage{amsfonts}
\usepackage{amsthm}
\usepackage{cases}
\usepackage{bm}
\usepackage{subcaption}

\usepackage{color}
\definecolor{Blue}{rgb}{0.00, 0.00, 1.00}
\definecolor{Red}{rgb}{1.00, 0.00, 0.00}
\definecolor{Green}{rgb}{0.00, 0.70, 0.00}

\hypersetup{
    colorlinks=true,       
    linkcolor=blue,          
    citecolor=blue,        
    filecolor=magenta,      
    urlcolor=cyan           
}

\begin{document}

\title{Two dimensional Coulomb gas in a non-conservative trap}
\author{David S. Dean}
\affiliation{Universit\'e Bordeaux, CNRS, LOMA, UMR 5798, F-33400 Talence, France}
\author{Rashed Aljasmi}
\affiliation{Center for Quantum and Topological Systems, NYUAD Research Institute, New York University Abu Dhabi, UAE}
\author{Satya N. Majumdar}
\affiliation{Laboratoire de Physique Th\'eorique et Mod\`eles Statistiques (LPTMS),
CNRS, Univ. Paris-Sud, Universite Paris-Saclay, 91405 Orsay, France}

\begin{abstract}

We study the nonequilibrium steady state of a two dimensional Coulomb gas under the 
action of an 
anisotropic harmonic trapping potential along with a non-conservative rotational force.  In 
the case without rotation, the equilibrium (zero temperature) steady state has a uniform 
density supported over a a static elliptical droplet. The addition of a rotational force 
drives the system into a nonequilibrium steady state where the density is still uniform 
inside an ellipse, but the ellipse gets tilted by a a fixed angle compared to the 
non-rotational case. In addition, a nonzero current is generated inside the droplet which 
run concentrically to the droplet boundary.
For large rotational force, the droplet develops 
a purely circular form. Our results are predicted by a simple hydrodynamic calculation and 
are confirmed by numerical simulations and provide a full understanding of a novel driven 
non-equilibrium state in a strongly interacting system.

\end{abstract}

\maketitle

\section{Introduction}
The one component two dimensional Coulomb gas is of considerable importance in statistical mechanics \cite{bru66,bau80,jan81,joh83,cun17,all14}, firstly as it is an example of a strongly interacting system that has soluble equilibrium statistical mechanics, but also because it is related to a number of random matrix ensembles \cite{met67,for10,for16}. Here we consider the zero temperature dynamics of a trapped ensemble of like charged particles  in a trap which has  a non-potential component at zero temperature. In equilibrium, when trapped by a conservative quadratic potential, the two dimensional Coulomb gas has, at large length scales, a uniform density and is in fact hyper-uniform \cite{tor18}. For a generic quadratic trapping potential  the positions of the gas particles have a one to one correspondence with the eigenvalues of random matrix theory. In the case of an isotropic trap one obtains the Ginibre ensemble \cite{gin65} corresponding to complex random matrices with independent and identically distributed real and imaginary coefficients (they are therefore not symmetric and do not have only real eigenvalues). The support of the eigenvalues of the Ginibre ensemble is a disc in the complex plane. The elliptic Ginibre \cite{gir86,ake23,byu25} ensemble however  has a broken rotational invariance and the support of the eigenvalues in this case is an ellipse. From the Coulomb gas point of view the difference between the Ginibre and elliptic Ginibre ensembles is that the confining potential is anisotropic while the pairwise interaction between the charges is always the two dimensional (logarithmic) Coulomb interaction. This anisotropy in the potential means that the confined gas is squeezed more along the axis where the harmonic trap is the stiffest, thus leading to the elliptical shape. 

In the mathematical literature there has been much study of the two dimensional Coulomb gas or $\beta$ ensemble where the Hamiltonian for the system is given over the complex numbers $z_k= x_k+i y_k$, corresponding to the position of particle $k$, as

\begin{equation}
H= \sum_{k=1}^N V(z_k) - \frac{q^2}{2\pi}\sum_{j<k} \ln(|z_j-z_k|),
\end{equation}
where $q$ is the charge of the particles. 
The case 
\begin{equation}
V(z) = |z|^2 -s Re(z^2) = x^2 (1-s) + y^2(1+s)
\end{equation}
for $s\in(0,1)$ describes the eigenvalues of the random normal matrix ensemble. The case $s=0$ and with $\beta= q^2/(2\pi k_BT )=2$ (where $T$ is  temperature and $k_B$ Boltzmann's constant), corresponds to the Ginibre ensemble when the harmonic confining potential is isotropic. The long-range nature of the 2d Coulomb interaction means that for large $N$, where $N$ is the number of particles, thermal fluctuations, and thus entropy,  are not important  
and the system of particles enters a frozen steady state which minimizes the total energy of the system. Numerically this can be simulated by using a simple gradient descent dynamics. This dynamics corresponds to Langevin dynamics without the noise term (hence the term zero temperature). The final steady state is locally given by the so called Wigner crystal structure, however on scales bigger than the inter-particle separation the density of charge is uniform. In this paper we consider the case where the trap confining the Coulomb gas generates a non-potential force. In general the overall force can be Helmholtz decomposed into a purely potential, irrotational,  quadratic one along with a purely rotational, or solenoidal, term. 
In the absence of the rotation component the steady state consists of a static elliptical droplet. However when a rotation term is introduced the steady state is modified in two ways. First the form of the droplet changes with respect to the irrotational case, the droplet is rotated and deformed to become more circular and less elliptic. Secondly, the overall form of the droplet, once in the steady state and as in the irrotational case, does not rotate or deform. However, the particles inside the droplet move along lines concentric to the overall droplet boundary to form concentric rotating lines of current. Another example of a non-equilibrium steady state has been investigated in \cite{pak18}; here the authors considered the Hamiltonian dynamics of two dimensional vortices (interacting again via the 2d Coulomb potential). In this setting, the system relaxes to steady state exhibiting a core-halo structure \cite{pak18}. The core is a vortex dense region which is elliptical and the halo a less dense region enclosing the core. Here, the region containing the vortices rotates, in contrast to the model studied here where the bounding region is fixed but with internal currents. 

The paper is organized as follows. In the next section \ref{model} we explain our model and notably the dynamics of the system. We  study the system in the hydrodynamic limit in section (\ref{hydro}). This approach is essentially exact for this system at large system sizes and in terms of the trap, rotation parameters, charge and system size we show that the steady state density support has  an elliptical form, which we predict analytically, and when a rotation component is applied the elliptical support deforms and currents are generated. Again, analytical expressions are given for all of these quantities. In section (\ref{num}) we carry out simple numerical simulations by integrating the equations of motion for a finite number of particles. These simulations confirm our analytical predictions.

\section{The sheared Coulomb gas model }
\label{model}
We consider a system of $N$ identical charges of  charge $q$ which are subject to their own 2d Coulomb interaction but also an external force  ${\bf F}_t$ due to a trap. We assume that the dynamics is  over-damped and zero temperature, the equation of motion of the particles with positions denoted by ${\bf x}_i$ and having friction coefficients $\gamma$ is  thus given by 
\begin{equation}
\gamma\frac{d{\bf x}_i}{dt}  = -q\nabla\phi({\bf x}_i)+ {\bf F}_t({\bf x}),
\end{equation}
where $\phi({\bf x})$ is the electrostatic potential obeying Poisson's equation
\begin{equation}
\nabla^2 \phi({\bf x}) = -q\rho({\bf x}),
\end{equation}
where
\begin{equation}
\rho({\bf x})=\sum_{i=1}^N \delta({\bf x}-{\bf x}_i),
\end{equation}
denotes  the particle number  density.

Now, if  we set $\gamma=1$ and $q=1$, then the  over-damped equation of motion for each particle $i$ among a total of $N$ particles is given by
\begin{equation}
\frac{d {\bf x}_i}{dt} = {\bf u}({\bf x}_i) -\nabla\phi({\bf x}_i),
\end{equation}
where ${\bf u}$ is the contribution to the drift coming from the trapping potential
\begin{equation}
{\bf u}({\bf x})=  {\bf F}_t({\bf x}).
\end{equation}
In two dimensions the  electrostatic potential is  given by
\begin{equation}
\phi({\bf x}) = -\frac{1}{2\pi}\sum_{i}\ln(|{\bf x}-{\bf x}_i|),
\end{equation}
and the over-damped noiseless equation of motion that we will study is therefore
\begin{equation}
\frac{d {\bf x}_i}{dt} = {\bf u}({\bf x}_i)+\frac{1}{2\pi}\sum_{j\neq i} \frac{{\bf x}_i-{\bf x}_j}{|{\bf x}_i-{\bf x}_j|^2}.\label{em}
\end{equation}
In what follows, we will consider the case where the flow field ${\bf u}$ takes the linear form, which is the general case anisotropic harmonic trapping potentials in the presence of linear non-conservative forces:
\begin{equation}
{\bf u} = U{\bf x},
\label{sp.1}
\end{equation}
and where, in the appropriately rotated frame of reference,  $U$ has a potential (corresponding to its symmetric part) and nonconservative components (corresponding to its antisymmetric part)
\begin{equation}
\label{U_def}
U =-\begin{pmatrix} & \mu_1 & 0\\ & 0 &\mu_2 \end{pmatrix} + \omega \begin{pmatrix} & 0 & 1\\ & -1 &0. \end{pmatrix}.
\end{equation}
We note that for some optical trapping set ups \cite{grier08,grier09,grier10,grier15,man19,ama19} there is indeed a non-conservative component to the trapping potential due to the radiation pressure.

To gain some intutions as to what one may expect in this system, it is instructive to
first study a single particle subject to the nonconservative force as in Eq. (\ref{sp.1})
with $U$ given in (\ref{U_def}), i.e., when one switches off the repulsive interaction.
In this case, we simply have
\begin{eqnarray}
\label{sp.2}
\frac{dx}{dt} &= & - \mu_1\, x+ \omega\, y \\
\frac{dy}{dt} &=& - \mu_2\, y- \omega\, x \, ,
\end{eqnarray} 
starting from some arbitrary initial condition, e.g., $x(0)=x_0>0$ and $y(0)=0$.
This deterministic linear equation can be trivially solved and one finds that
the nature of the solution depends on whether $|\mu_1-\mu_2|<2 \omega$ or
$|\mu_1-\mu_2|>2 \omega$. In the former case, the solution reads
\begin{eqnarray}
\label{sp_sol.1}
x(t)&= & x_0\, e^{-(\mu_1+\mu_2) t/2}\left[ \frac{\mu_2-\mu_1}{\Omega}\, \sin\left(
\frac{\Omega t}{2}\right) + \cos\left(\frac{\Omega t}{2}\right)\right] \\
y(t)&=& - \frac{2\omega x_0}{\Omega}\, e^{-(\mu_1+\mu_2) t/2}\, \sin\left(\frac{\Omega t}{2}\right)\, ,
\end{eqnarray}
where $\Omega= \sqrt{4 \omega^2 - (\mu_1-\mu_2)^2}>0$. Thus in this case, the position
of the particle in the $(x,y)$ plane approaches to the origin exponentially fast as expected,
but it moves in a spiral way, with a clockwise rotation starting from $(x_0, 0)$.
In the case when $|\mu_1-\mu_2|>2 \omega$, there is no spiralling solution but
$(x(t), y(t))$ moves initially to $y<0$ (starting from $x_0>0$ and $y=0$) 
and then decays exponentially fast to the origin. Thus the role of the rotation force
with a nonzero $\omega>0$ is to provide a {\em torque} on the  the particle, forcing it 
to rotate clockwise.

We now turn to what  happens if one now switches on the repulsive interaction between the particles in
the presence of a nonzero $\omega>0$. It will be  
shown in the next section that this drives the system into a nonequilibrium steady state 
where the average density of the particles is supported uniformly over an ellipse in the 
$(x,y)$ plane, with $a$ and $b$ denoting respectively its semi-major and semi-minor axes 
lengths. However, this ellipse gets tilted by an angle $\theta$ (compared to the case 
$\omega=0$) clockwise. This clockwise rotation of the elliptical support can be intuitively 
guessed from the single particle picture above, but the angle $\theta$ by which it rotates, 
as well as the axes lengths $a$ and $b$ are fixed by the interaction between the particles. 
Our goal in the next section is to compute exactly, using a macroscopic hydrodynamic 
approach, these three unknowns $(a, b, \theta)$ determining the shape and the
orientation of the droplet 
supporting the average density. In addition, we will show that there are nonzero currents 
circulating inside the droplet and we will compute the average current density in the 
nonequilibrium steady state.

\section{Hydrodynamic study}\label{hydro}

In this section we will use a hydrodynamic approach to study the steady state properties of 
the Coulomb gas in two dimensions subject to a non-potential trapping force. In the case 
where the force is purely harmonic the steady state can be obtained by minimizing the energy 
of the system and it is well known that the support of the charges is elliptical 
\cite{ake23,byu25}. However the hydrodynamic approach proposed here allows us to study both 
the case where the force has a nonconservative component. We first show that we can find a 
solution where the hydrodynamic density is constant in the general case. In the potential 
case we show that the steady state hydrodynamic equation admits a solution with zero current 
(as it is an equilibrium system) and where the density has an elliptical support. The 
elliptical support is then computed for this $\omega=0$ case as shown in Fig. 
(\ref{fig410}). In the general case, $\omega\neq 0$ we compute the form of the confining 
ellipse as well as the tilting of the ellipse due to the nonconservative rotational force 
with respect to the orientation of the ellipse in the conservative case. One can see for 
example how the principal axes of the equilibrium ellipse in Fig. (\ref{fig410}) 
($\omega=0$) becomes tilted in Fig. (\ref{fig411}) and Fig. (\ref{fig415}) when $\omega=1$ 
and $\omega=5$ respectively. Finally we see in Fig. (\ref{fig411}) and Fig. (\ref{fig415}) 
that when $\omega\neq 0$, the particles have a non-zero velocity due to the current which is 
present, we will compute this current ${\bf j}$ and then use this to derive the local 
particle velocity ${\bf v}({\bf x})$ as a function of position in the non-equilibrium steady 
state.

In addition to computing the form of the ellipse we will also derive the tilt angle $\theta$.
The equation for the evolution of the density field $\rho({\bf x})$ is 
\begin{equation}
\frac{\partial \rho({\bf x})}{\partial t}=-\nabla\cdot {\bf j}({\bf x}), \label{eqfull}
\end{equation}
where
\begin{equation}
{\bf j}({\bf x}) = \rho({\bf x})[{\bf u}({\bf x}) -\nabla\phi({\bf x})].
\end{equation}
This hydrodynamic transport equation can also be derived from first principle
via stochastic density functional theory \cite{kaw94,dea96} with temperature set to zero.

The stability of a single particle (and thus the system) requires that $\mu_1, \mu_2>0$. 
Notice that the random normal matrix ensemble corresponds to the case $\mu_1= 2+2t$ , 
$\mu_2=2-2t$ and $\omega=0$.

In the next section, from  numerical simulations we will see that the system appears to have a uniform density spread over an elliptical support. When $\omega\neq 0$ the particles move to form currents which run along the edges of  ellipses which are concentric to the bounding ellipse. 

To proceed, for large $N$, we assume that in the steady state and within the support of the density, $\rho({\bf x})$ can be treated as continuous and constant giving the steady state equation:
\begin{equation}
\nabla\cdot{\bf j}({\bf x}) = \rho\nabla\cdot[{\bf u}({\bf x}) -\nabla\phi({\bf x})]=0.\label{steady}
\end{equation}
We comment here that the assumption of a constant density is inspired by the numerical simulations carried out in the following section (\ref{num}), however we will see that this assumption does lead to a consistent solution of the problem which agrees with the numerical simulations. It would be interesting to provide a proof of uniqueness of the solution of Eq. (\ref{eqfull}) in the long time limit.
The steady state  equation (\ref{steady}) then  gives 
\begin{equation}
\nabla^2\phi({\bf x}) +\mu_1+\mu_2 = 0 
\end{equation}
and so from Poisson's equation 
\begin{equation}
\rho = \mu_1+\mu_2,
\end{equation}
showing the consistency of the assumption that $\rho({\bf x})=\rho$ is constant. Notice that in the case where the ${\bf u}({\bf x})$ is derived from a potential one can rigorously prove that $\rho$ must be constant.  Now, we assume that the steady state solution is spread over an ellipse which when rotated so that its principal axes are aligned with the cartesian axes $(x,y)$ has the equation
\begin{equation}
\Phi({\bf x}) = \frac{x^2}{a^2} + \frac{y^2}{b^2} =1.\label{ell}
\end{equation}
The electrostatic potential for a disc with uniform charge $\rho$ is given, up to a constant,  by \cite{kel53,pak18}
\begin{equation}
\phi({\bf x}) = -\frac{\rho a b}{2 (a+b)}\left( \frac{x^2}{a} + \frac{y^2}{b}\right).
\label{pot}
\end{equation}
In terms of the particle number $N$ one then has
\begin{equation}
\phi({\bf x}) = -\frac{N}{2\pi(a+b)}\left( \frac{x^2}{a} + \frac{y^2}{b}\right).
\end{equation}
The steady state equation clearly has the solution 
\begin{equation}
{\bf j}({\bf x})= {\bf 0},
\end{equation}
when there is a solution for $\phi$ to the equation
\begin{equation}
{\bf u}({\bf x}) -\nabla\phi({\bf x})={\bf 0},
\end{equation}
that is to say when ${\bf u}({\bf x}) $ is purely conservative.
When this is not the case, the solution is given by
\begin{equation}
\nabla\cdot {\bf j}({\bf x}) = 0.
\end{equation}
If the bounding surface has the form $\Phi({\bf x})= {\rm constant}$ then there must be  no flux through the bounding surface and the concentric surfaces of the ellipse, therefore
\begin{equation}
\nabla\Phi({\bf x}) \cdot {\bf j}({\bf x}) = 0,
\end{equation}
as $\nabla\Phi({\bf x})$ is parallel to the normal vector of the ellipse.
This then gives
\begin{equation}
\nabla\Phi({\bf x})\cdot[{\bf u}({\bf x}) -\nabla\phi({\bf x})]=0.
\end{equation}
We now write 
\begin{equation}
\Phi({\bf x})= \frac{1}{2} {\bf x}\cdot B {\bf x},
\end{equation}
and
\begin{equation}
\phi({\bf x})= -\frac{1}{2} {\bf x}\cdot C {\bf x}.
\end{equation}
This then gives the equation
\begin{equation}
{\bf x}\cdot B(U+C){\bf x}=0.
\end{equation}
This means that the matrix $A = B(U+C)$ is antisymmetric and thus 
\begin{equation}
B(U+C)+(U^T + C)B =0,\label{mat}
\end{equation}
where we have used the fact that $C=C^T$ and $B=B^T$. The steady state current is therefore given by
\begin{equation}
{\bf j}({\bf x}) = \rho(U+C){\bf x}.
\end{equation}

In the frame where the axes of the ellipse are aligned with the Cartesian axes (so after a rotation by an angle $\theta$) both the $B$ and $C$ are simultaneously diagonalised. We have
\begin{equation}
R(\theta) BR^T(\theta) = B_d\ {\rm and }\ R(\theta) CR^T(\theta)= C_d,
\end{equation}
where $R(\theta)$ is the rotation matrix
\begin{equation}
R(\theta) = \begin{pmatrix} & \cos(\theta) & -\sin(\theta)\\& \sin(\theta) & \cos(\theta) \end{pmatrix},
\end{equation}
and from Eq. (\ref{ell}) we obtain
\begin{equation}
B_d=2\begin{pmatrix} &\frac{1}{a^2}& 0\\ &0 &\frac{1}{b^2}\end{pmatrix},
\end{equation}
and from Eq. (\ref{pot}) we find
\begin{equation}
C_d=\frac{N}{\pi(a+b)}\begin{pmatrix} &\frac{1}{a}& 0\\ &0 &\frac{1}{b}\end{pmatrix}.
\end{equation}
Now transforming Eq. (\ref{mat}) as
\begin{equation}
R(\theta)[B(U+C)+(U^T + C)B]R^T(\theta) =0,
\end{equation}
we find
\begin{equation}
B_dR(\theta)UR^T(\theta)+ R(\theta)U^TR^T(\theta)B_d + 2 B_d C_d=0,\label{me2}
\end{equation}
where we have used the fact that $B_d$ and $C_d$ are both diagonal and so commute. Then, Eq. 
(\ref{me2}) leads to three equations which can be written as
\begin{eqnarray}
\mu_1 \cos^2(\theta)+\mu_2 \sin^2(\theta) &=&\frac{N}{\pi a(a+b)}\\
\mu_2 \cos^2(\theta)+\mu_1 \sin^2(\theta) &=&\frac{N}{\pi b(a+b)}\\
(\mu_1-\mu_2)\sin(2\theta) &=& \frac{2\omega(b^2-a^2) }{(a^2+ b^2)}.
\label{three_eq.1}
\end{eqnarray}
Summing the first two equations gives
\begin{equation}
\mu_1+ \mu_2 = \rho = \frac{N}{\pi ab}.\label{murho}
\end{equation}
One can scale out the dependence on $N$ by writing $a=\alpha\sqrt{N}$ and $b=\beta\sqrt{N}$ in which case the equations become
\begin{eqnarray}
\mu_1 \cos^2(\theta)+\mu_2 \sin^2(\theta) &=&\frac{1}{\pi \alpha(\alpha+\beta)}\label{a1}\\
\mu_2 \cos^2(\theta)+\mu_1 \sin^2(\theta) &=&\frac{1}{\pi \beta(\alpha+\beta)}\label{a2}\\
(\mu_1-\mu_2)\sin(2\theta) &=& \frac{2\omega(\beta^2-\alpha^2) }{(\alpha^2+ \beta^2)}\label{a3}.
\end{eqnarray}
Thus, we have three equations for the three unknown parmaeters $(\alpha,\beta, \theta)$.
The pair $(\alpha, \beta)$ characterizes the shape of the ellipse, 
while $\theta$ fixes the orientation of the ellipse with respect to the $\omega=0$ case.
In terms of the rescaled lengths $(\alpha, \beta)$, Eq. (\ref{murho}) reduces to
\begin{equation}
\mu_1+ \mu_2 = \frac{1}{\pi \alpha\beta}.
\end{equation}

By eliminating $(\alpha, \beta)$ from Eq. (\ref{a3}), one obtains
a single nonlinear equation for $\theta$ (when $\mu_1\neq \mu_2$):
\begin{equation}
 \sin(2\theta) =\frac{ 8\omega\cos(2\theta)(\mu_1+\mu_2)}{3\mu_1^2 + 2 \mu_1\mu_2 + 3 \mu_2^2 + (\mu_1-\mu_2)^2\cos(4\theta)},\label{theta}
\end{equation}
and with the ellipse axes given in terms of $\theta$ by
\begin{eqnarray}
\alpha &=&\sqrt{\frac{1}{\pi(\mu_1+\mu_2)}\frac{\mu_1\sin^2(\theta) + \mu_2\cos^2(\theta)}{\mu_1\cos^2(\theta) + \mu_2\sin^2(\theta)}}\label{ax1n}\\
\beta &=&\sqrt{\frac{1}{\pi(\mu_1+\mu_2)}\frac{\mu_2\sin^2(\theta) + \mu_1\cos^2(\theta)}{\mu_2\cos^2(\theta) + \mu_1\sin^2(\theta)}}.\label{ax2n}
\end{eqnarray}
The equation for $\theta$ can be solved numerically (it can actually be written as a cubic equation in $\tan(2\theta)$ which can be solved but the expression is very complicated). 

There are in fact three cases where the above set of equations have  simple solutions 

\noindent {\bf Case 1: $\omega=0$.} This is the equilibrium case and we see that $\theta=0$ and $\theta=\pi$ are the two, obviously physically equivalent,  solutions. The ellipse then has axes
\begin{eqnarray}
\alpha&=& \frac{\sqrt{\frac{\mu_2}{\mu_1}}}{\sqrt{\pi}\sqrt{\mu_1+\mu_2}}\label{ax1}\\
\beta &=& \frac{\sqrt{\frac{\mu_1}{\mu_2}}}{\sqrt{\pi}\sqrt{\mu_1+\mu_2}}.\label{ax2}
\end{eqnarray}
One should note here that the anisotropic harmonic potential acting on the charges due to  the trap  can in fact be thought of as being generated by a uniform distribution of background charge density $- \rho$ on the ellipse given by Eq. (\ref{ell}) (so a jellium model). The charges in the trap thus distribute uniformly to neutralize the background charge as shown in \cite{forr96}. 

\noindent {\bf Case 2: $\omega\to\infty$.} 
Note that as $\omega \to \infty$ one see that the solution for $\theta$ must be given by $\cos(2\theta) = 0$ and so $\theta=\pm\pi/4$. However, one sees that when $\omega >0$ but is finite, the corresponding solution is such that  $\theta>0$.  Consequently, by continuity,  as $\omega$ increases we expect $\theta$ to tend to the positive solution $\pi/4$. However the solutions $\theta=\pm\pi/4$ are indeed physically equivalent as 
in the large $\omega$ limit we obtain a circular support with
\begin{equation}
\alpha =\beta = \sqrt{\frac{1}{\pi(\mu_1+\mu_2)}}.
\end{equation}
It is particularly interesting that strong rotation leads to a circular fixed point.

\noindent {\bf Case 3: $\mu_1=\mu_2$.} Here the solution is simply
\begin{equation}
\theta=0; \ \alpha=\beta = \frac{1}{\sqrt{2\pi\mu_1}},
\end{equation}
however this solution does have a current. The current here is rather trivial though and one can show that if one goes to a rotating frame with angular velocity $\omega$, then the dynamical equations Eq. (\ref{em}) become the equilibrium ones for the case $\omega=0$. In fact in the case  where $\mu_1=\mu_2$, we see from the  equations (\ref{a1}, \ref{a2}, \ref{a3}) that all values of $\theta$ are  solutions by due to the rotational invariance of the system. However Eq. (\ref{theta}) is valid in the case where $\mu_2\neq \mu_1$ but without equality. Therefore  Eq. (\ref{theta}) is valid for an almost circular ellipse $\mu_1\to\mu_2$. Here we find that Eq. (\ref{theta}) becomes
\begin{equation}
\tan(2\theta) =\frac{ 2\omega}{\mu_1},
\end{equation}
and so
\begin{equation}
\theta = \frac{1}{2}\tan^{-1}{\frac{ 2\omega}{\mu_1}}.
\end{equation}
Again we see that  when $\omega\to\infty$ one finds that $\theta=\pi/4$.

After determining the tilt angle $\theta$ and the form of the elliptical support we can now compute the current of the non-equilibrium steady state:
\begin{equation}
{\bf j}({\bf x}) = J{\bf x}\label{j},
\end{equation}
where 
\begin{equation}
J = \rho[U + R^T(\theta) C_d R(\theta)].
\end{equation}
The above then gives, again in terms of the angle $\theta$ and using Eq. (\ref{murho}), 
\begin{equation}
J = \frac{1}{2}(\mu_2^2-\mu_1^2)\sin(2\theta)\begin{pmatrix}& \sin(2\theta)&  \cos(2\theta)\\
& \cos(2\theta) & -\sin(2\theta)\end{pmatrix} +(\mu_1+\mu_2) \begin{pmatrix}& 0& \omega \\
& -\omega & 0\end{pmatrix}.\label{J}
\end{equation}
Also, note that as $\nabla\cdot {\bf j}({\bf x})=0$ we have ${\rm Tr}(J)=0$. In the isotropic case where  $\mu_1=\mu_2$, when the support is circular, the current its simply given by the non-conservative part of the flow field. In the limit $\omega\to\infty$ we see that 
\begin{equation}
J = (\mu_1+\mu_2)\begin{pmatrix}& 0&  \omega \\
& -\omega & 0\end{pmatrix}.
\end{equation}
We note that as the density is constant, we can define a local velocity vector for the velocity of the particles as a function of their position via ${\bf j}({\bf x}) = \rho {\bf v}$ and so from Eq. (\ref{murho}) and Eq. (\ref{J}) we find 
\begin{equation}
{\bf v}=  [\frac{1}{2}(\mu_2-\mu_1)\sin(2\theta)\begin{pmatrix}& \sin(2\theta)&  \cos(2\theta)\\
& \cos(2\theta) & -\sin(2\theta)\end{pmatrix} + \begin{pmatrix}& 0& \omega \\
& -\omega & 0\end{pmatrix}]{\bf x}.\label{V}
\end{equation}

We have thus seen that in the absence of a rotation component in the trapping force or due to shear, the charges relax to a uniform density, bounded by an ellipse with axes given by Eqs (\ref{ax1},\ref{ax2}),  the charges are frozen and the ellipse has no internal dynamics. The introduction of  a rotation component leads to the original ellipse being tilted or rotated by an angle $\theta$ given by Eq. (\ref{theta}) and a modification of the axes, given by Eqs. (\ref{ax1n},\ref{ax2n}), which depend on the degree of rotation $\theta$. At large values of the rotation component $\omega$, the overall rotation of the ellipse saturates at $\theta=\pm\pi/4$ and the ellipse becomes a circle. Finally, while in the presence of a rotational component, the steady state form of the bounding ellipse is static,  a current is however induced inside the ellipse and is given by the linear form Eq. (\ref{j}) with $J$ given by Eq. (\ref{J}). As the rotation component $\omega \to\infty$ the ellipse is deformed into a circle with concentric circular currents within.
\section{Numerical Simulations}\label{num}
\begin{figure}
    \includegraphics[width=10cm,height=10cm]{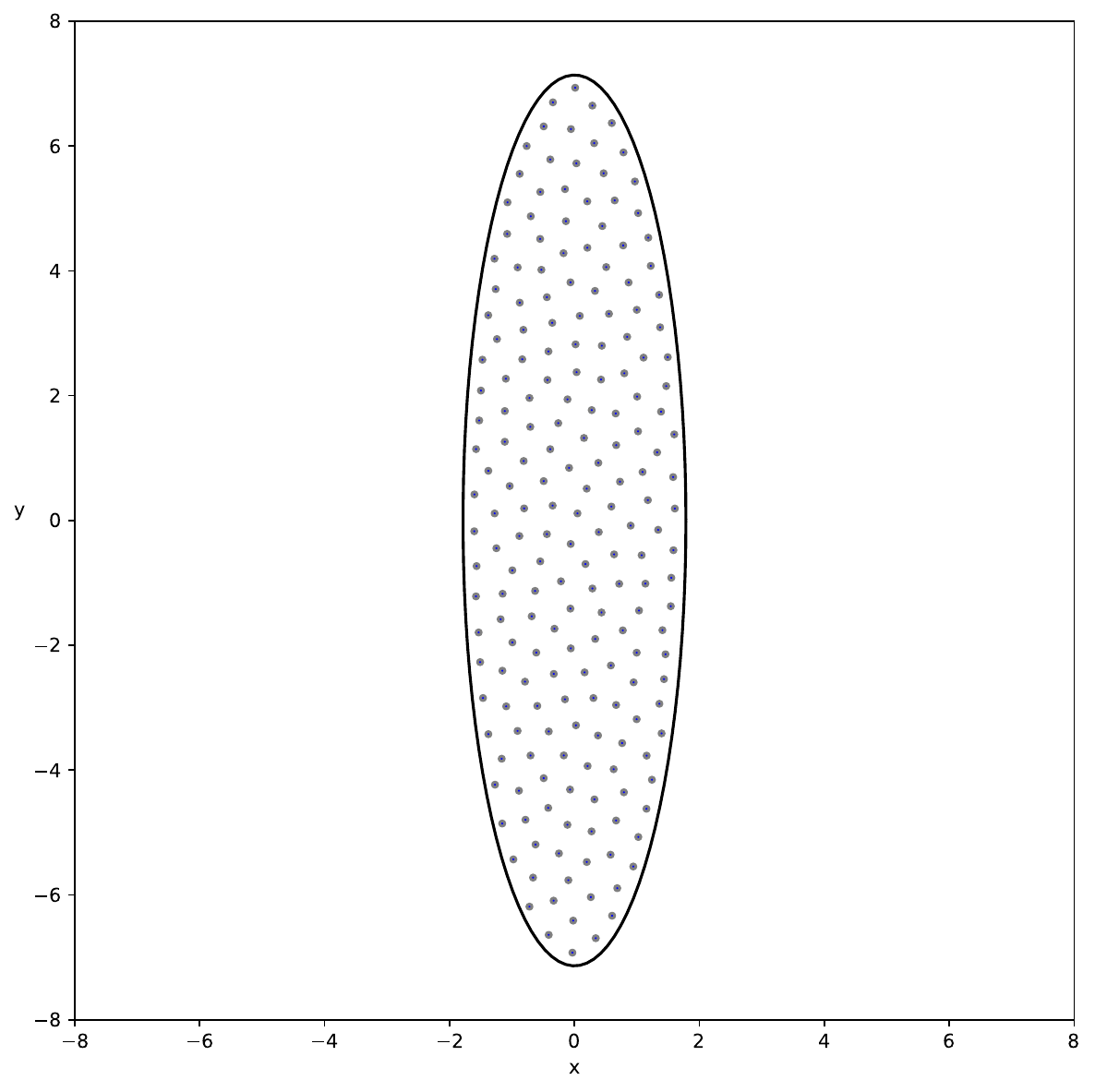}
    \caption{Numerical simulation for $N=200$ particles (blue dots) with $\mu_1=4$, $\mu_2=1$ and $\omega=0$. Shown in black is the enclosing ellipse as predicted by the theory. The particles are stationary.}  \label{fig410}
\end{figure}

The numerical simulations are performed by integrating the equations of motion (\ref{em}) with a simple forward Euler scheme. We start with a distribution of charges where each is placed uniformly at random on the region $[-1/2,1/2]\times [-1/2,1/2]$. 
From this initial configuration, the system then expands and reaches a steady state. In the case of no rotation component, $\omega=0$, the charges are indeed found to be  bounded by an ellipse and the distribution of charges is uniform on length scales larger than the average inter-particle separation. Shown in Fig. (\ref{fig410}) is the simulation result for $N=200$ particles (grey circles) for $\mu_1=4$ and $\mu_2=1$ but $\omega=0$ corresponding to the conservative case. We see that the black bounding ellipse predicted by the theory (here $\theta=0$) is in perfect agreement with the simulation results. At small length scales we see a locally deformed Wigner type triangular lattice.

\begin{figure}
    \includegraphics[width=10cm,height=10cm]{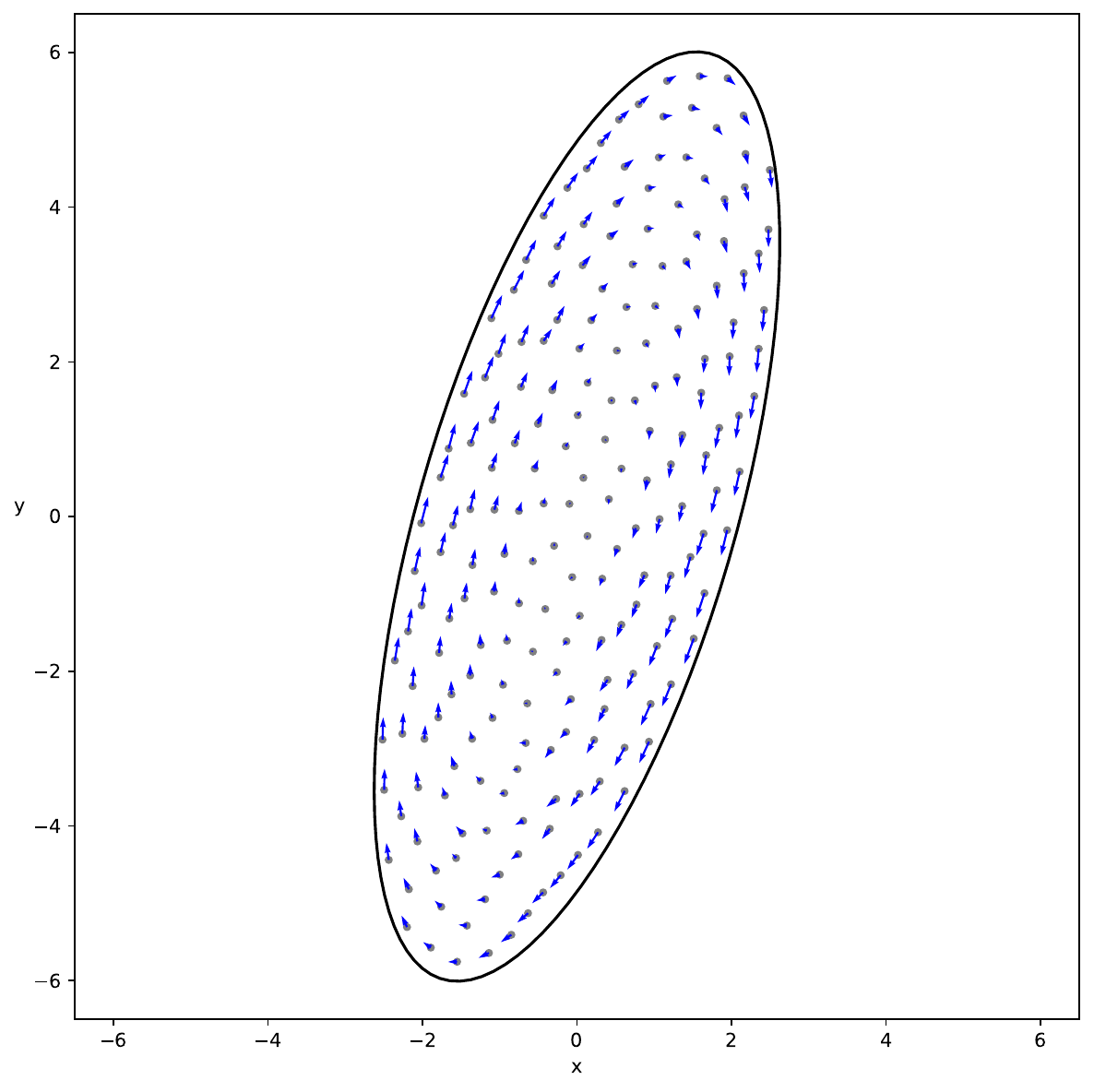}
    \caption{Numerical simulation for $N=200$ particles (blue dots) with $\mu_1=4$, $\mu_2=1$ and $\omega=1$. Shown in black is the enclosing ellipse as predicted by the theory. The arrows on the particles indicate (up to the same rescaling for each particle for visibility) the particles' velocity vector.}  \label{fig411}
\end{figure}

\begin{figure}
    \includegraphics[width=10cm,height=10cm]{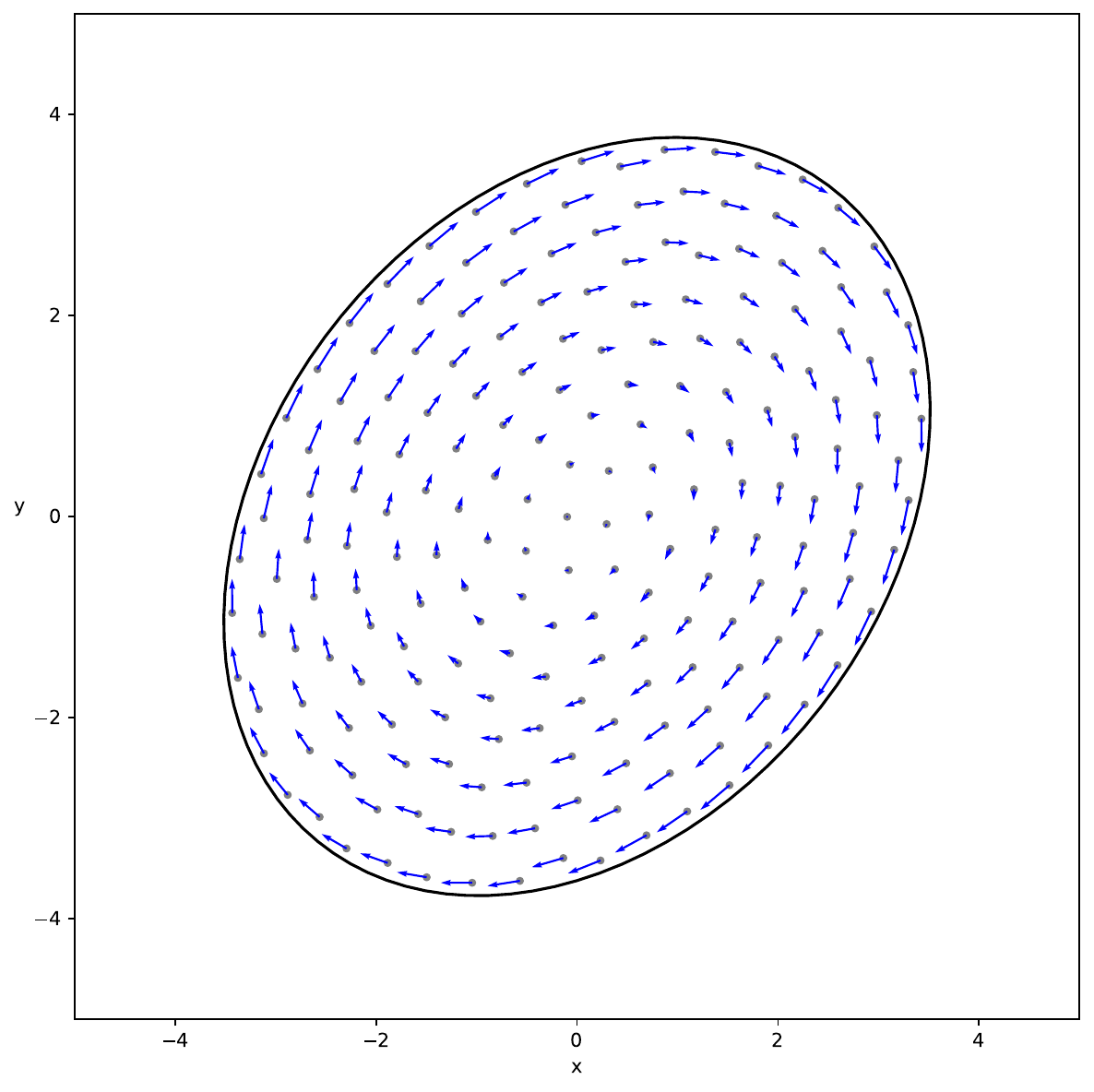}
    \caption{Numerical simulation for $N=200$ particles (blue dots) with $\mu_1=4$, $\mu_2=1$ and $\omega=5$. Shown in black is the enclosing ellipse as predicted by the theory. The arrows on the particles indicate (up to the same rescaling for each particle for visibility) the particles' velocity vector.}  \label{fig415}
\end{figure}

Keeping the same parameters for $\mu_1$ and $\mu_2$ but taking $\omega=1$, theory predicts  that $\theta = 0.28355$. Note this means that the supporting ellipse when rotated by the angle $\theta$ has its axes aligned with the Cartesian coordinate axes and so the ellipse observed in simulations is rotated clockwise by the angle $-\theta$. The resulting predicted confining ellipse agrees perfectly with the theoretical prediction, as shown in Fig. (\ref{fig411}). Interestingly, the particles do not follow perfectly concentric ellipses; instead, kinks or defects are observed. Increasing the rotation to $\omega=5$ we find $\theta = 0.66203$, again the agreement with theory shown in Fig. (\ref{fig415}) matches theoretical predictions. In this more circular geometry the type of kinks seen in the case $\omega=1$ become rarer.

\begin{figure}[ht]
  \centering
  \begin{subfigure}[b]{.5\textwidth}
    \centering
    \includegraphics[width=\linewidth]{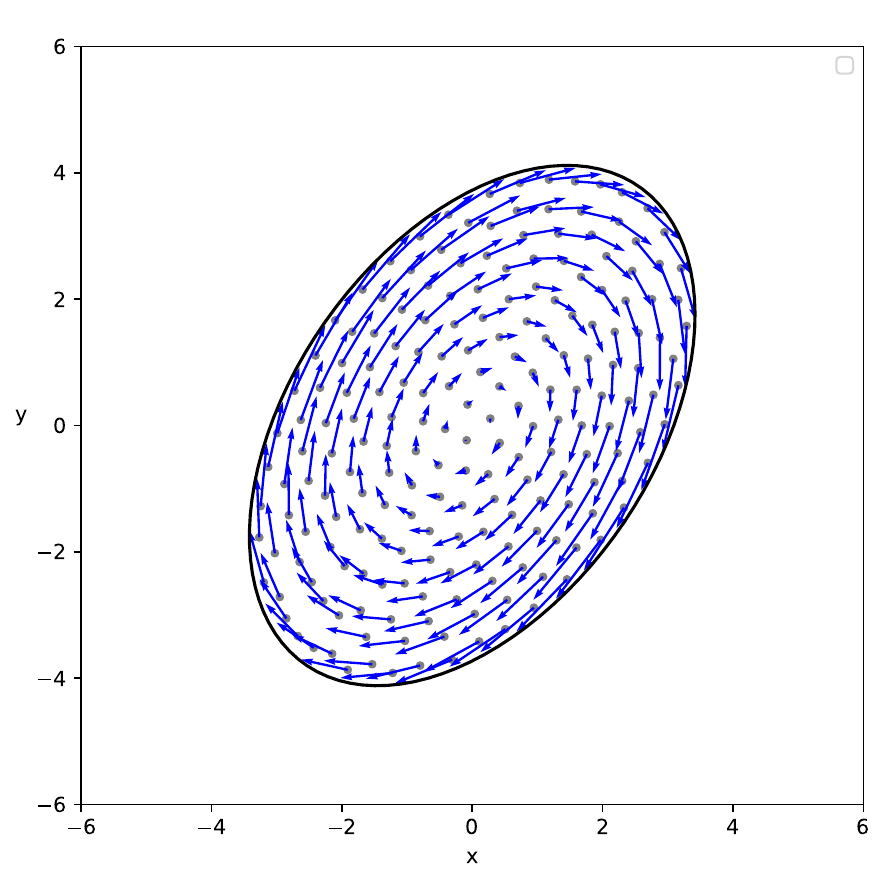}
    \caption{}
  \end{subfigure}
  \begin{subfigure}[b]{.5\textwidth}
    \centering
    \includegraphics[width=\linewidth]{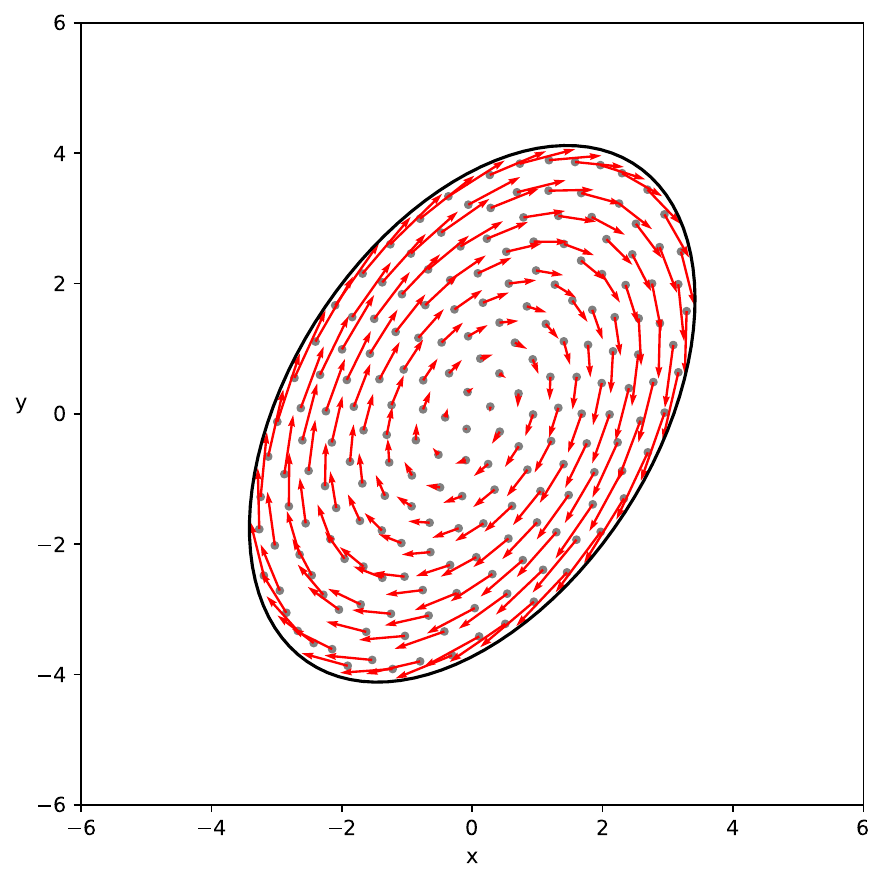}
    \caption{}
  \end{subfigure}
   \caption{\label{Vfig}(a) The difference in the measured steady velocities (scaled by $1/10$ for clarity) and (b) the theoretically predicted value from Eq. (\ref{V}) for a system of 200 particles for $\mu_1=4$, $\mu_2=1$ and $\omega=5$. (also scaled by $1/10$ for clarity). The difference is practically not visible when plotted on the same scale.}
\end{figure}

Another analytical prediction is for the current given by Eq. (\ref{J}), however given the fact that the analytical calculation is hydrodynamic in nature, it is more practical to use the prediction for the velocity of the particles ${\bf V}$ given in Eq. (\ref{V}). In Fig. (\ref{Vfig}a) is we show the numerical simulations for $200$ particles, $\mu_1=4$, $\mu_2=1$ and $\omega=3$ (the rotation angle being $\theta=0.577685$) with the velocity vectors shown for each particle  rescaled by a factor of $1/10$ for clarity. In Fig. (\ref{Vfig}b) we show the same particle positions but with the velocity vector ${\bf v}({\bf x}_i)$ for each particle position ${\bf x}_i$ predicted from Eq. (\ref{V}), again rescaled by a factor of $1/10$ for clarity. If one computes the difference between the measured and predicted velocities and plots them on the same graph with the same scaling the differences are not visible, showing the high accuracy of the hydrodynamic predictions.

\section{Conclusions}
We have studied the steady state of a two dimensional Coulomb gas which is subjected to an external force corresponding to the superposition of an isotropic harmonic potential and an additional rotational term not generated by a potential.  The case with no rotational term has been extensively studied, especially for an isotropic harmonic potential as this corresponds to the so called jellium model \cite{bru66,bau80,jan81,joh83}. The case of an anisotropic harmonic trap has been less studied;  however it corresponds to the eigenvalues of the Ginibre ensemble of random matrices in the complex plane \cite{met67,for10,for16,ake23,byu25}. At length scales above the inter-particle separation, a hydrodynamic analysis predicts that, as in the conservative case, the density is uniform and the support of the density is an ellipse. The effect of a rotational term is to tilt the ellipse and generate internal currents within the bounding ellipse, which itself remains stationary. These internal currents are predicted to be made up of concentric shells within the ellipse. We have seen that the analytical predictions are confirmed by numerical simulations for system sizes as small as 200 particles.. In the context of equilibrium systems, there have been very detailed studies of the finer structure of two dimensional Coulomb gases, including correlation functions, full counting statistics and extreme value statistics \cite{cun16a,cun16b,lac18}. The analytical results based on the hydrodynamic approach used here cannot be applied to these properties; however, numerical simulations suggest interesting features worth exploring, particularly the kinks in particle motion that represent deviations from elliptical concentric currents.

The Riesz gas presents a generalisation of the two dimensional Coulomb gas to long-range interacting models in arbitrary dimensions, where the interactions are pairwise and power law. For a trapped Riesz gas \cite{rie38,lew22} in a power law confining potential,  the distances and couplings can be rescaled in such a way that, in the large $N$ limit the equilibrium density can be determined by energy minimization. It would be interesting to see if this approach can be extended to develop a hydrodynamic analysis for  Riesz gases under non-conservative external forces. It would be particularly interesting to examine the case where the steady state density is not uniform leading to a more complex steady state configuration.

\begin{acknowledgments}
 D.S.D. and S.N.M acknowledge support  from the grant ANR Grant No. ANR-23-CE30-0020-01 EDIPS.
\end{acknowledgments}


\begin{thebibliography}{10}
\bibitem{bru66} S.G. Brush, H.L. Sahlin and E. Teller, Monte Carlo Study of a One‐Component Plasma. I, J. Chem. Phys. {\bf 45}, 2102 (1966).
\bibitem{bau80} M. Baus and J.-P. Hansen, Statistical Mechanics of simple Coulomb systems, Physics Reports {\bf 59}, 1, (1980).
\bibitem{jan81}B. Jancovici, On the two-dimensional Coulomb gas, Phys. Rev. Lett. {\bf 46}, 386 (1981).
\bibitem{joh83}S Johannesen and D Merlini, On the thermodynamics of the two-dimensional jellium, J. Phys. A: Math. Gen. {\bf 16}, 1449 (1983).
\bibitem{cun17}F.D. Cunden, P. Facchi, M. Ligab\`o, and   P. Vivo, Universality of the third-order phase transition in the constrained Coulomb gas,  J. Stat. Mech.  053303 (2017).
\bibitem{all14} R. Allez, J. Touboul, and G. Wainrib, Index distribution of the Ginibre ensemble,  J.  Phys. A: Mathematical and Theoretical {\bf 47} 042001 (2014).
\bibitem{met67}
M. L. Mehta, {\it Random Matrices}, Academic Press (1967).
\bibitem{for10}
P. J. Forrester, {\it Log-Gases and Random Matrices}, Princeton University Press (2010).
\bibitem{for16} P.J. Forrester, Analogies between random matrix ensembles and the one-component plasma in two-dimensions, Nucl. Phys.  B {\bf 904}, 253 (2016).
\bibitem{tor18} S. Torquato, {\em Hyperuniform states of matter}, Phys. Rep. {\bf 745}, 95 (2018)
\bibitem{gin65} J. Ginibre, Statistical ensembles of complex, quaternion and real random matrices, J. Math. Phys. {\bf 6}, 440 (1965). 
\bibitem{gir86}V. L. Girko, {\em Elliptic law}, Theory Probab. Appl. {\bf 30}, 677 (1986).
\bibitem{ake23} G. Akemann, M. Duits and L.D. Molag, The elliptic Ginibre ensemble: A unifying approach to local and global statistics for higher dimensions, J. Math. Phys. {\bf 64}, 023503 (2023)
\bibitem{byu25}
S.-S. Byun and P. J. Forrester, {\it Progress on the Study of the Ginibre Ensembles}, Springer Nature (2025).
\bibitem{pak18}R. Pakter and Y. Levin, Non-equilibrium Statistical Mechanics of Two-dimensional Vortices,
Phys. Rev. Lett. {\bf 121}, 020602 (2018).
\bibitem{grier08} Y. Roichman, B. Sun, A. Stolarski and D. G. Grier, Physical Review Letters {\bf 101}, 128301 (2008).
\bibitem{grier09}B. Sun, J. Lin, E. Darby, A. Y. Grosberg and D. G. Grier, Brownian vortexes, Physical Review E {\bf 80}, 010401(R) (2009).
\bibitem{grier10} B. Sun, D. G. Grier and A. Y. Grosberg, Minimal model for Brownian vortexes, Physical Review E {\bf 82}, 021123 (2010).
\bibitem{grier15} H.W. Moyses, R.O. Bauer, A.Y. Grosberg and D.G. Grier, Perturbative theory for Brownian vortexes, Phys. Rev. E {\bf 91}, 062144 (2015).
\bibitem{man19}M. Mangeat, Y. Amarouchene, Y. Louyer, T. Guérin and D. S. Dean, Role of non-conservative scattering forces and damping on  Brownian particles in optical traps, Phys. Rev. E {\bf 99}, 052107 (2019).
\bibitem{ama19} Y. Amarouchene,  M. Mangeat, B. Vidal Montes, L. Ondic, T. Guerin, D. S. Dean  and Y. Louyer,  Nonequilibrium dynamics induced by scattering forces for optically trapped nanoparticles in strongly inertial regimes, Phys. Rev. Lett. 122, 183901 (2019).		
\bibitem{kaw94} K. Kawasaki, Stochastic model of slow dynamics in supercooled liquids and dense colloidal suspensions, Physica A: Statistical Mechanics and its Applications {\bf 208}, 35 (1994).
\bibitem{dea96}D. S. Dean, Langevin equation for the density of a system of interacting Langevin processes, J. Phys. A: Math. Gen. {\bf 29}, L613 (1996).
\bibitem{kel53}O. D. Kellogg, Foundations of Potential Theory (Dover,
New York, 1953).
\bibitem{forr96} P. J. Forrester and  B. Jancovici, Two-dimensional one-component plasma in a quadrupolar field, Int. J.
Mod. Phys. A {\bf 11} 941 (1996).
\bibitem{cun16a}  F. D. Cunden, F. Mezzadri and Pierpaolo Vivo,Large deviations of radial statistics in the two-dimensional one-component plasma,J. Stat. Phys. (2016) {\bf 164}, 1062 (2016).
\bibitem{cun16b}  F. D. Cunden, P. Facchi, P. Vivo, A shortcut through the Coulomb gas method for spectral linear statistics on random matrices, J. Phys. A: Math. Theor. {\bf 49}, 135202 (2016).
\bibitem{lac18}
B. Lacroix-A-Chez-Toine, A, Grabsch, S. N. Majumdar, and G. Schehr, Extremes of 2d Coulomb gas: universal intermediate deviation regime, J. Stat. Mech.  013203, (2018).
\bibitem{rie38}M. Riesz, Riemann Liouville integrals and potentials., Acta Sci. Math. Univ. Szeged {\bf 9}, 1 (1938).
\bibitem{lew22} M. Lewin, Coulomb and Riesz gases: The known and the unknown,  J.  Math. Phys.  {\bf 63},  (2022).








\end{thebibliography}
\end{document}